\documentclass[final,5p,times,twocolumn]{elsarticle}
\usepackage{graphicx}
\usepackage{bm}
\usepackage{amsmath}
\usepackage{amsfonts}
\usepackage{amssymb}
\biboptions{compress}
\newcommand {\be} {\begin{equation}}
\newcommand {\ba} {\begin{eqnarray}}
\newcommand {\ee} {\end{equation}}
\newcommand {\ea} {\end{eqnarray}}
\journal{Physics Letters B}

\begin{document}

\begin{frontmatter}

\title{Total hadronic cross sections at high energies in holographic QCD}

\author[IHEP,UCAS]{Akira Watanabe}

\ead{akira@ihep.ac.cn}

\author[IHEP,UCAS]{Mei Huang}

\ead{huangm@ihep.ac.cn}

\address[IHEP]{Institute of High Energy Physics and Theoretical Physics Center for Science Facilities, Chinese Academy of Sciences, Beijing 100049, People's Republic of China}

\address[UCAS]{University of Chinese Academy of Sciences, Beijing 100049, People's Republic of China}

%%%%%%%%%%%%%%%%%%%%%%%%%%%%%%%%%%%%%%%%
\begin{abstract}
We present our analysis on high energy hadron-hadron scattering in the framework of the holographic QCD.
Combining the Brower-Polchinski-Strassler-Tan Pomeron exchange kernel and gravitational form factors of the involved hadrons which are obtained by using the bottom-up AdS/QCD models in the five-dimensional AdS space, we calculate the total cross sections at high energies.
We show that our calculations for the nucleon-nucleon scattering agree with the experimental data including the recent ones taken by the TOTEM collaboration at the LHC.
The present framework is applicable to any high energy process, in which the strong interaction can be approximated by the Pomeron exchange.
We present the results for the pion-nucleon and pion-pion scattering as examples, which can be obtained without any additional parameter because all the adjustable parameters are fixed via the analysis on the nucleon-nucleon scattering.
The resulting total cross section ratios are $\sigma_{tot}^{\pi N} / \sigma_{tot}^{N N} = 0.63$ and $\sigma_{tot}^{\pi \pi} / \sigma_{tot}^{N N} = 0.45$, respectively.
\end{abstract}
%%%%%%%%%%%%%%%%%%%%%%%%%%%%%%%%%%%%%%%%

\begin{keyword}
Gauge/string correspondence \sep AdS/CFT correspondence \sep Pomeron \sep LHC
\end{keyword}

\end{frontmatter}

%%%%%%%%%%%%%%%%%%%%%%%%%%%%%%%%%%%%%%%%
\section{\label{sec:level1}Introduction}
%%%%%%%%%%%%%%%%%%%%%%%%%%%%%%%%%%%%%%%%
Experimentally measurable high energy scattering processes have played important roles in revealing the internal structures of hadrons and the quark-gluon dynamics inside them over several decades.
Recently, the data taken at LHC provide us great opportunities to deepen our knowledge about the Standard Model, especially about the strong interaction.
Focusing on QCD, cross sections of high energy processes can be factorized into the soft and hard parts, according to the factorization theorem.
The hard part can be calculated in principle by using the perturbation technique in QCD.
However, on the other hand, the soft part cannot be rigorously calculated because of its nonperturbative nature, and the only way to treat it is the phenomenological parameterization of the parton distribution functions (PDFs).
PDFs are expressed with two kinematic variables, the Bjorken scaling variable $x$ and the energy scale $Q^2$.
Since the recent improvements in experimental techniques enable us to access the higher energy regime, a solid understanding of the small Bjorken $x$ physics is certainly required.
Due to this background, improving nonperturbative methods to describe QCD in such a kinematic region has been one of the most important subjects in hadron physics.

The holographic QCD, which is constructed based on the AdS/CFT correspondence~\cite{Maldacena:1997re,Gubser:1998bc,Witten:1998qj}, is one of the effective approaches to QCD, and has gathered a lot of theoretical interests so far~\cite{Kruczenski:2003be,Son:2003et,Kruczenski:2003uq,Sakai:2004cn,Erlich:2005qh,Sakai:2005yt,DaRold:2005zs}.
Since Polchinski and Strassler first applied the AdS/CFT correspondence to the analysis on high energy scattering~\cite{Polchinski:2001tt}, many related studies have been done.
In particular, the deep inelastic scattering (DIS) has been intensively studied~\cite{Polchinski:2002jw,BoschiFilho:2005yh,Brower:2006ea,Hatta:2007he,Brower:2007qh,BallonBayona:2007rs,Brower:2007xg,Cornalba:2008sp,Pire:2008zf,Cornalba:2010vk,Brower:2010wf,Watanabe:2012uc,Watanabe:2013spa,Watanabe:2015mia}, and understanding its small $x$ dynamics is especially important as mentioned above.
To describe the complicated partonic dynamics in the small $x$ region, assuming the Pomeron exchange, which can be interpreted as a multi-gluon exchange, is useful~\cite{Donnachie:1992ny,ForshawRoss,PomeronPhysicsandQCD}.
Brower, Polchinski, Strassler, and Tan (BPST) performed the gauge/string duality based analysis in the higher dimensional AdS space, and proposed a kernel which gives a Pomeron exchange contribution to the cross sections in high energy scattering phenomena~\cite{Brower:2006ea}.
The BPST kernel was further improved subsequently~\cite{Brower:2007qh,Brower:2007xg}, and applied to the analysis on structure functions in DIS at small $x$~\cite{Brower:2010wf,Watanabe:2012uc,Watanabe:2013spa,Watanabe:2015mia}.
The most striking result may be that the nontrivial scale dependence of the Pomeron intercept, which were experimentally measured at HERA~\cite{Breitweg:1998dz}, was well reproduced with this kernel.

In this work, we apply the BPST kernel to the analysis on the total hadronic cross sections of hadron-hadron scattering at high energies.
Compared to DIS, hadron-hadron scattering seems to be a simpler process, but this is not necessarily correct in the theoretical study in particular.
In DIS, correctly producing the $Q^2$ dependence, where $Q^2$ is the four-momentum squared of the probe photon, may be most important, but the description of the photon is established in both the QCD and string theory, which relies on the fact that the photon is an elementary particle.
On the other hand, in hadron-hadron scattering, since the involved scales are the masses of the participants only, the descriptions of the involved hadrons, which are nonperturbative composites, become more important.
Hence, describing the hadron-hadron scattering is also challenging.
Recently, new data for the proton-proton total cross sections at the TeV scale have been provided by the TOTEM collaboration at LHC~\cite{Antchev:2017dia}.
The energy 13~TeV, at which the latest data were taken, corresponds to $x \sim 10^{-8}$, considering the proton mass $m_p \sim 1$~GeV.
This $x$ value is much smaller compared to the $x$ range which was investigated at HERA.
Therefore, it is expected that testing a phenomenological model in this kinematic region may provide us a new insight to deepen our understanding of the Pomeron nature.

In our model setup, we combine the BPST Pomeron exchange kernel and two density distributions of the involved hadrons in the five-dimensional AdS space, and calculate the total cross sections.
Since the Pomeron exchange can be realized as the Reggeized graviton exchange in the AdS space, the density distributions are described by the gravitational form factors, which are obtained by using the bottom-up AdS/QCD models~\cite{Abidin:2008hn,Abidin:2008ku,Abidin:2009hr}.
It is known that the hard-wall model, in which the AdS geometry is sharply cut off in the infrared (IR) region to introduce the QCD scale, can give a reasonable description unless we consider an excited state of the hadron, so we employ the hard-wall models to describe the hadrons in this study.
It is shown that both the recent TOTEM data and other pp and $\bar{\rm{p}}$p data can be well reproduced within our model.
In the framework, once the adjustable model parameters are fixed with the data, one can predict the hadron-hadron cross sections involving other hadrons without any additional parameter, because a hadron is a normalizable mode and its density distribution is normalized.
As examples, besides the nucleon-nucleon case, we also present the pion-nucleon and pion-pion results, which shows the versatility of our model and may support further applications to other high energy scattering processes in which the Pomeron exchange gives a reasonable approximation.

In the next section, we explain the model setup, and then present our numerical results in Sec.~\ref{sec:level3}.
The summary and discussion are given in Sec.~\ref{sec:level4}.

%%%%%%%%%%%%%%%%%%%%%%%%%%%%%%%%%%%%%%%%
\section{\label{sec:level2}Holographic description of hadron-hadron total cross sections}
%%%%%%%%%%%%%%%%%%%%%%%%%%%%%%%%%%%%%%%%
Here we present our model setup to obtain the total cross sections of hadron-hadron scattering, employing the BPST Pomeron exchange kernel denoted by $\chi$ in the equations hereafter~\cite{Brower:2006ea}.
Following the preceding study~\cite{Brower:2010wf}, the scattering amplitude of the two-body process, $1 + 2 \to 3 + 4$, in the five-dimensional AdS space can be expressed in the eikonal representation as
\be
{\cal A} (s, t) = 2 i s \int d^2 b e^{i \bm{k_\perp } \cdot \bm{b}} \int dzdz' P_{13}(z) P_{24}(z') \left[ 1-e^{i \chi (s, \bm{b}, z, z')} \right], \label{eq:amplitude}
\ee
where $s$ and $t$ are the Mandelstam variables, $\bm{b}$ is the two-dimensional impact parameter, and $z$ and $z'$ are fifth (or bulk) coordinates for the incident and target particles, respectively.
$P_{13}(z)$ and $P_{24}(z')$ in Eq.~\eqref{eq:amplitude} represent the density distributions of the two hadrons in the AdS space, which are to be specified for the numerical evaluations, and satisfy the normalizable conditions,
\be
\int dz P_{13}(z) = \int dz' P_{24}(z') = 1, \label{eq:normalization_condition}
\ee
because the both particles are normalizable modes in this study.

Utilizing the optical theorem and picking up the leading contribution to the eikonal approximation in Eq.~\eqref{eq:amplitude}, which means that we consider the single-Pomeron exchange, the total cross section is expressed as
\be
\sigma_{tot} (s) = 2 \int d^2b \int dzdz' P_{13} (z) P_{24} (z') \mbox{Im} \chi (s,\bm{b},z,z'). \label{eq:tcs_original}
\ee
In the conformal limit, the analytical expression of the imaginary part of $\chi$ can be obtained, and the impact parameter integration in Eq.~\eqref{eq:tcs_original} can be performed analytically.
Hence, Eq.~\eqref{eq:tcs_original} is rewritten as
\begin{align}
&\sigma_{tot} (s) = \frac{g_0^2 \rho^{3/2}}{8 \sqrt{\pi}} \int dzdz' P_{13} (z) P_{24}(z') (zz') \mbox{Im} [\chi_{c}(s,z,z')], \label{eq:tcs_with_CK} \\
&\mbox{Im} [\chi_c(s,z,z') ] \equiv e^{(1-\rho)\tau} e^ {-[({\log ^2 z/z'})/{\rho \tau}]} / {\tau^{1/2}}, \label{eq:CK}
\end{align}
where
\be
\tau = \log (\rho z z' s / 2).
\ee
Here $g_0^2$ and $\rho$ are adjustable parameters which control the magnitude and the energy dependence of the cross sections, respectively.

It is known that the inclusion of the confinement effect is necessary to reproduce the experimental data of the structure functions measured at HERA with the BPST Pomeron exchange kernel, unless we focus on the hard scattering, i.e., the high $Q^2$ region in the DIS case~\cite{Brower:2010wf,Watanabe:2012uc,Watanabe:2013spa}.
Considering the nucleon-nucleon scattering, the scale of the process can be characterized by the nucleon mass, $m_N \sim 1$~GeV, which implies that the low energy dynamics with a strong coupling in QCD becomes dominant.
Hence, instead of the conformal kernel, in this study we utilize the modified BPST kernel, in which the confinement effect is mimicked, given with the same functional form as the conformal kernel by
\begin{align}
&\mbox{Im} [\chi_{mod} (s, z, z')] \equiv 
\mbox{Im} [\chi_c (s, z, z') ] \nonumber \\
&\hspace{27mm} + \mathcal{F} (s, z, z') \mbox{Im} [\chi_c (s, z, z_0 z_0' / z') ],\label{eq:MK} \\
&\mathcal{F} (s, z, z') = 1 - 2 \sqrt{\rho \pi \tau} e^{\eta^2} \mbox{erfc}( \eta ), \\
&\eta = \left( -\log \frac{z z'}{z_0 z_0'} + \rho \tau \right) / {\sqrt{\rho \tau}},
\end{align}
where $z_0$ and $z'_0$ are the cutoffs of the fifth coordinates which characterize the QCD scale.
Note that these two parameters are uniquely fixed with hadron masses, so not adjustable.

To perform the numerical evaluations of the total cross sections, one needs to specify the density distributions, $P_{13}(z)$ and $P_{24}(z')$, which characterize the involved hadrons.
These distributions are expressed by the gravitational form factors which can be extracted from the hadron-Pomeron-hadron three point functions, using the bottom-up AdS/QCD models of hadrons.
Since we consider the nucleon-nucleon, pion-nucleon, and pion-pion scattering in this study, we need to specify the density distributions of the nucleon and the pion.
The gravitational form factors of the both hadrons were obtained by the authors of Refs.~\cite{Abidin:2008hn,Abidin:2009hr}, so we can utilize those results.
In the model discussed in Refs.~\cite{Henningson:1998cd,Muck:1998iz,Contino:2004vy,Hong:2006ta}, the nucleon is described as a solution to the five-dimensional Dirac equation.
Following the previous study~\cite{Watanabe:2012uc}, the density distribution of the nucleon is expressed with the left-handed and right-handed components of the Dirac field, $\psi_L$ and $\psi_R$, respectively in terms of the Bessel function as
\begin{align}
&P_N (z) = \frac{1}{2z^{3}}  \left[ \psi_L^2 (z) + \psi_R^2 (z) \right], \\
&\psi_L (z) = \frac{\sqrt{2} z^2 J_2 (m_N z)}{z_0^N J_2 (m_N z_0^N)}, \
\psi_R (z) = \frac{\sqrt{2} z^2 J_1 (m_N z)}{z_0^N J_2 (m_N z_0^N)},
\end{align}
where the cutoff parameter $z_0^N$ is fixed by the condition, $J_1 (m_N z_0^N) = 0$.
The value used in this study is $z_0^N = 1 / (245$~MeV$)$ with $m_N = (m_p + m_n)/2$, where $m_p$ and $m_n$ are the proton and neutron physical masses, respectively.
On the other hand, focusing on the chiral limit case, the pion wave function $\Psi$ can also be analytically obtained as a solution to the equation of motion derived from the bottom-up AdS/QCD model of mesons~\cite{Erlich:2005qh}.
Following the procedure in Ref.~\cite{Watanabe:2012uc} again, the density distribution of the pion is given with the Bessel function by
\begin{align}
&P_\pi (z) = \frac{ \left[ \partial _{z} \Psi (z) \right] ^2 }{4 \pi^2 f_\pi ^2 z}   + \frac{\sigma^2 z^6 \Psi (z)^2 }{ f_\pi ^2 z^3 }, \\
&\Psi \left( z \right) = z\Gamma \left[ {\frac{2}{3}} \right] \left( {\frac{\alpha }{2}} \right)^{1/3} \Biggl[ I_{ - 1/3} \left( {\alpha z^3 } \right) - I_{1/3} \left( {\alpha z^3 } \right)\frac{{I_{2/3} \left( {\alpha (z_0^\pi)^3 } \right)}}{{I_{ - 2/3} \left( {\alpha (z_0^\pi)^3 } \right)}} \Biggr] ,
\end{align}
where $f_\pi$ is the pion decay constant, $\alpha = 2 \pi \sigma / 3$, $\sigma = (332$~MeV$)^3$, and the cutoff parameter $z_0^\pi$ is fixed with the $\rho$ meson mass $m_\rho$ by the condition, $J_0 (m_\rho z_0^\pi) = 0$.
The actual value used in this work is $z_0^\pi = 1/(322$~MeV$)$.

%%%%%%%%%%%%%%%%%%%%%%%%%%%%%%%%%%%%%%%%
\section{\label{sec:level3}Numerical results}
%%%%%%%%%%%%%%%%%%%%%%%%%%%%%%%%%%%%%%%%
As seen in the previous section, since the expressions for the density distributions do not include any adjustable parameter, there are only two parameters in total, $g_0^2$ and $\rho$, in the BPST Pomeron exchange kernel, to be determined with experimental data.
Focusing on the energy range, $10^2 < \sqrt{s} < 10^5$~GeV, and considering the recent proton-proton collision data measured by the TOTEM collaboration at LHC~\cite{Antchev:2017dia,Antchev:2013gaa,Antchev:2013iaa,Antchev:2013paa,Antchev:2015zza,Antchev:2016vpy,Nemes:2017gut} and other pp~\cite{Baltrusaitis:1984ka,Honda:1992kv,Collaboration:2012wt} and $\bar{\rm{p}}$p~\cite{Battiston:1982su,Hodges:1983oba,Bozzo:1984rk,Alner:1986iy,Amos:1991bp,Abe:1993xy,Augier:1994jn,Avila:2002bp} data which were summarized by the Particle Data Group in 2010~\cite{Nakamura:2010zzi}, we perform the numerical fitting.
The best fit values we obtained are $g_0^2 = 6.27 \times 10^2$ and $\rho = 0.824$.
It should be mentioned here that $g_0^2$ is just an overall factor and the slope of the total cross section is totally governed by the other parameter $\rho$.
Also, the values we obtained in this work are different from those in Ref.~\cite{Watanabe:2012uc}, which is because the kinematics of hadron-hadron scattering is different from that of DIS.

\begin{figure*}[tb!]
\begin{center}
\includegraphics[width=0.75\textwidth]{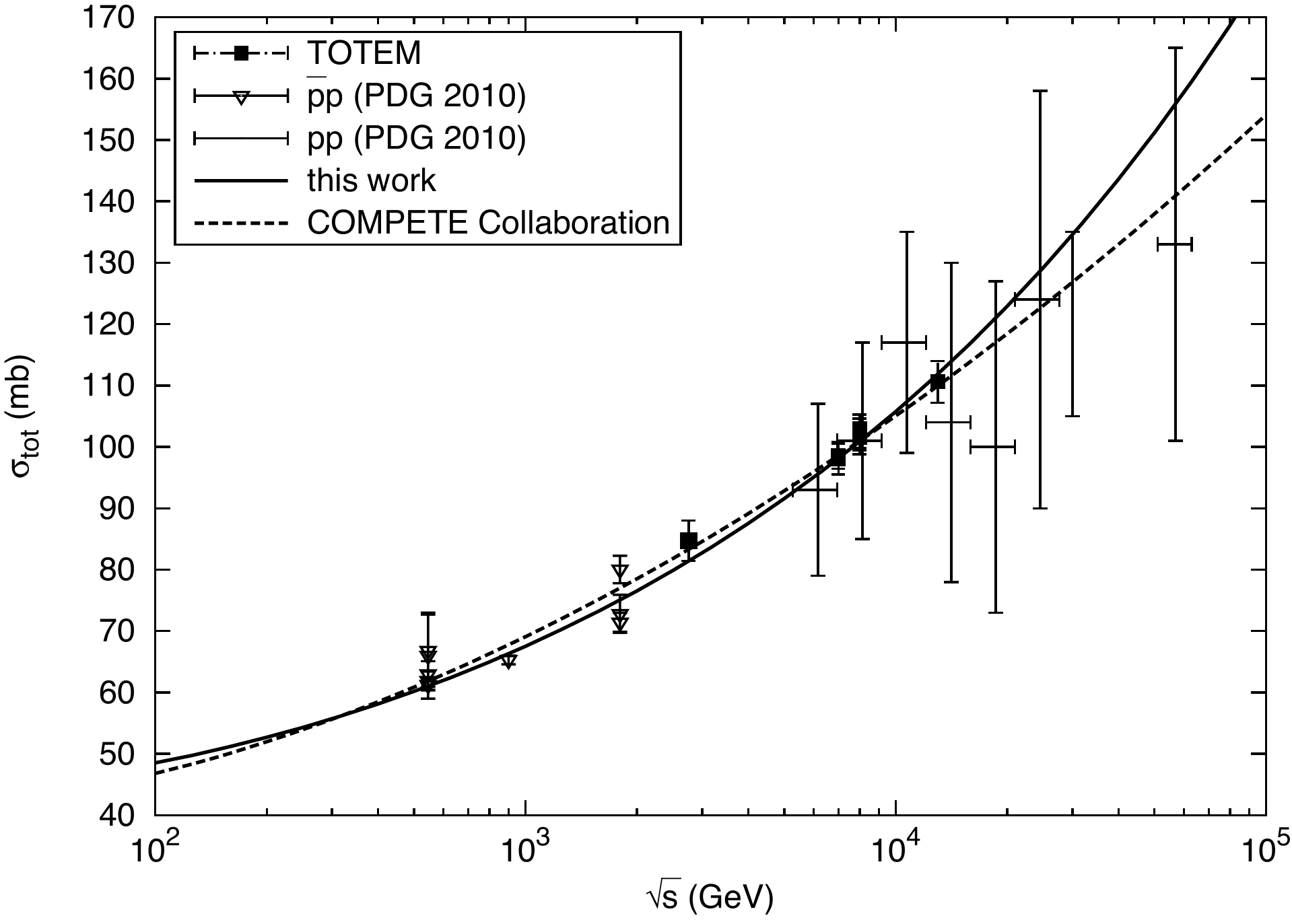}
\caption{
The nucleon-nucleon total cross section as a function of $\sqrt{s}$.
The solid and dashed curves represent our calculation and the empirical fit by COMPETE collaboration~\cite{Cudell:2002xe}, respectively.
The experimental data measured by the TOTEM collaboration at LHC~\cite{Antchev:2017dia,Antchev:2013gaa,Antchev:2013iaa,Antchev:2013paa,Antchev:2015zza,Antchev:2016vpy,Nemes:2017gut} and other data for pp~\cite{Baltrusaitis:1984ka,Honda:1992kv,Collaboration:2012wt} and $\bar{\rm{p}}$p~\cite{Battiston:1982su,Hodges:1983oba,Bozzo:1984rk,Alner:1986iy,Amos:1991bp,Abe:1993xy,Augier:1994jn,Avila:2002bp} collisions are depicted with error bars.
}
\label{fig:TCS_NN}
\end{center}
\end{figure*}

We display in Fig.~\ref{fig:TCS_NN}
the resulting nucleon-nucleon total cross section, compared with the experimental data and the empirical fit by COMPETE collaboration~\cite{Cudell:2002xe}.
One can see from the figure that our calculation is in agreement with the TOTEM data and other pp and $\bar{\rm{p}}$p data, although the data at $\sqrt{s} > 10$~TeV except for the TOTEM's have huge uncertainties because they were extracted from the cosmic-ray experiments.
Also, it can be seen that our result is consistent with the empirical fit at $\sqrt{s} < 10$~TeV, but a substantial deviation is observed in the higher $s$ region.

\begin{figure}[tb!]
\includegraphics[width=0.47\textwidth]{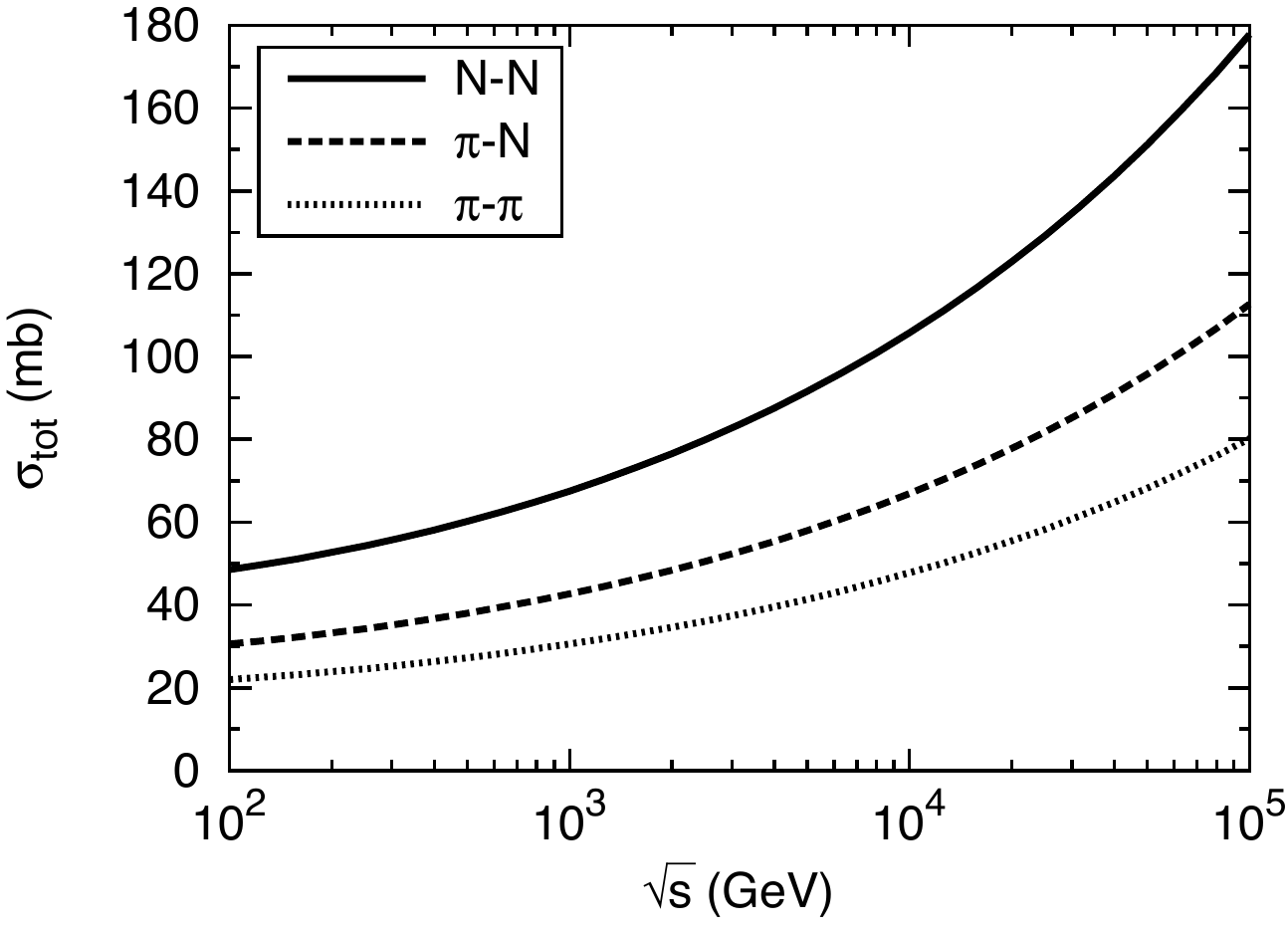}
\caption{
The resulting total cross sections as a function of $\sqrt{s}$.
The solid, dashed, and dotted curves represent our calculations for the nucleon-nucleon, pion-nucleon, and pion-pion scattering, respectively.
}
\label{fig:TCS_summary}
\end{figure}

Next, we show our results for the total cross sections of the pion-nucleon and pion-pion scattering in Fig.~\ref{fig:TCS_summary}.
The Pomeron exchange is a universal picture, and the adjustable parameters included in the BPST kernel do not depend on the involved hadron properties.
Also, in our present model setup both the nucleon and the pion are normalizable modes and follow the normalization conditions, Eq.~\eqref{eq:normalization_condition}.
Hence, once we fix the model parameters with the nucleon-nucleon data, we can predict the pion involved processes without any additional parameter.
Since the scales appeared in the considered scattering processes are characterized by the involved hadron masses, which are constants, the $s$ dependence is common to all the three curves in Fig.~\ref{fig:TCS_summary}.
Therefore, the total cross section ratios which can be obtained from the results are also constant.
Our predictions are as follows:
\be
\frac{\sigma_{tot}^{\pi N}}{\sigma_{tot}^{N N}} = 0.63, \
\frac{\sigma_{tot}^{\pi \pi}}{\sigma_{tot}^{N N}} = 0.45.
\ee
Donnachie and Landshoff studied various hadron-hadron total cross sections, considering the so-called soft Pomeron with the intercept 1.0808~\cite{Donnachie:1992ny}, and the ratio extracted from their results is $\sigma_{tot}^{\pi N} / \sigma_{tot}^{N N} = 0.63$ which agrees with our prediction.

%%%%%%%%%%%%%%%%%%%%%%%%%%%%%%%%%%%%%%%%
\section{\label{sec:level4}Summary and discussion}
%%%%%%%%%%%%%%%%%%%%%%%%%%%%%%%%%%%%%%%%
In this work, we have investigated the total cross sections of hadron-hadron scattering at high energies in the framework of holographic QCD, assuming the Pomeron exchange to describe the involved strong interaction.
The cross sections are expressed by combining the BPST Pomeron exchange kernel and the two density distributions of the involved hadrons.
The density distributions are described by the gravitational form factors which can be obtained from the bottom-up AdS/QCD models.
The comparison between our calculations and experimental data, including the recent ones measured by the TOTEM collaboration at LHC, has been explicitly demonstrated.

The resulting nucleon-nucleon total cross section is in agreement with the experimental data in the considered kinematic range, $10^2 < \sqrt{s} < 10^5$~GeV, which shows that our model works well in the high energy hadron-hadron scattering in addition to the previously studied DIS processes.
In DIS, the density distribution of the probe photon is concentrated around the ultraviolet (UV) boundary (small $z$), and the peak position of the target hadron density distribution is located in the IR region (large $z$).
This UV-IR scattering realizes the larger Pomeron intercept depending on the probe photon virtuality $Q^2$, compared to the soft Pomeron with a smaller constant intercept.
On the other hand, since the hadron-hadron scattering considered in this work is the IR-IR scattering, its kinematics is different from that of DIS.
Therefore, the observation that the total cross section data can be well reproduced within the model implies its wide applicability.

Our nucleon-nucleon result is also consistent with the empirical fit proposed by the COMPETE collaboration at $\sqrt{s} < 10$~TeV.
However, a substantial deviation is observed in the higher $s$ regime.
One possible reason for this may be a lack of the saturation effect in our model.
Although the interaction in the hadron-hadron scattering is soft, the gluonic saturation may occur in the TeV scale.
We have not taken into account this effect in the model so far, but the present observation implies the importance of considering this.
Our calculation of the nucleon-nucleon scattering is larger than that the empirical fit shows, which seems reasonable, because the saturation effect may suppress the cross section.

Besides the nucleon-nucleon scattering, we have studied the pion-nucleon and pion-pion scattering also.
Since the energy dependence of the cross section in our model setup is totally governed by the BPST kernel, the observed $s$ dependencies are common to all the results.
There is no data for such pion involved processes in the high $s$ region at this moment, but the predicted total cross section ratio $\sigma_{tot}^{\pi N} / \sigma_{tot}^{N N}$ agrees with the value extracted from the soft Pomeron based study done by Donnachie and Landshoff.
To pin down this, experimental data are expected to be taken with a high intensity pion beam in the future.

The results we obtained in this work suggest that the BPST Pomeron exchange kernel can be a useful and powerful theoretical tool in the analysis on various high energy scattering processes in which the nonperturbative gluonic dynamics can be approximated by the Pomeron exchange.
One of the important extensions may be to consider differential cross sections within the present framework.
The TOTEM collaboration has measured the proton-proton differential cross sections, and those data would be useful to further constrain the model parameters.
Moreover, applications to other processes, such as the deeply virtual Compton scattering or the vector meson production, may also be interesting.
Further studies are certainly needed.

%%%%%%%%%%%%%%%%%%%%%%%%%%%%%%%%%%%%%%%%
\section*{Acknowledgements}
%%%%%%%%%%%%%%%%%%%%%%%%%%%%%%%%%%%%%%%%
A.W. acknowledges Marco Ruggieri for fruitful discussions.
This work was supported in part by the NSFC under Grant Nos. 11725523, 11735007, and 11261130311 (CRC 110 by DFG and NSFC).

\bibliographystyle{elsarticle-num}

\section*{References}

\bibliography{hQCD}

\end{document}